%% file: richardsa_sept04.tex
\documentclass{evn2004}
\usepackage{txfonts}
\usepackage{graphicx}
\begin{document}
\include{page}
   \title{Hidden bipolarity in red supergiant winds}

   \author{A.M.S. Richards\inst{1}
          \and
          M.R.W. Masheder\inst{2}
\and H.J. van Langevelde\inst{3}
\and R.J. Cohen\inst{1}
\and M.D. Gray\inst{4}
\and J.A. Yates\inst{5}
\and P.J. Diamond\inst{1}
\and W. H. T. Vlemmings\inst{6}
\and M. Szymczak\inst{7}
\and K. Murakawa\inst{8}
          }

   \institute{Jodrell Bank Observatory, University of Manchester, Macclesfield,
        Cheshire, SK11, UK.
              \email{amsr@jb.man.ac.uk}
         \and
             Department of Physics, University of Bristol, Tyndall Avenue, Bristol, BS8 1TL, UK.
\and Joint Institute for VLBI in Europe, Postbus 2, 7990 AA Dwingeloo, The Netherlands.
\and  Department of
Physics, UMIST, PO Box 88, Manchester M60 1QD, UK.
\and Department of Physics and
Astronomy, University College London, Gower Street, London WC1E 6BT,
UK.
\and Department of Astronomy, Cornell University, 610 Space Sciences Building, Ithaca, NY 14853, USA.
\and Toru\'{n} Centre for Astronomy, Nicolaus Copernicus University, ul. Gagarina 11, PL-87100 Toru\'{n}, Poland.
\and ASTRON,  Postbus 2, 7990 AA Dwingeloo, The Netherlands.
             }

   \abstract{Many observations of late-type M stars show large
   near-spherical circumstellar envelopes, yet planetary nebulae and
   supernova remnants are frequently axisymmetric.  We present VLBI
   and MERLIN observations of masers around the red supergiant S Per
   which show varying degrees of axisymmetry and a dynamically
   significant magnetic field.  There is no evidence for rotation here
   or in most similar objects.  We examine possible origins of the magnetic field.  }

   \maketitle
%

\section{Introduction}
\label{md}


Red supergiants (RSG) with progenitor masses $\ga10$~M$_{\odot}$ are
high-mass analogues of Asymptotic Giant Branch (AGB) stars such as
Mira variables, typically one or a few M$_{\odot}$.  Both classes of
star have surface temperatures of 2000 -- 3000 K and pulsational
periods of up to a few years.  SiO, H$_{2}$O and OH masers are found
in the winds and show that the outflow velocities are similar to
within a factor of two. Such properties appear to be determined by
small-scale physics, such as the opacity of the stellar atmosphere or
the absorptivity of dust, and hence depend primarily on the local
chemical composition.

Other phenomena do scale with the stellar mass.  RSG have stellar radii
of $R_{\star}\ga5$ au and mass loss rates
$dM/dt\approx10^{-5}-10^{-4}$, at least ten times higher than those
of AGB stars. For all such stars the H$_2$O maser zone extends from
$\sim 5 - 30R_{\star}$ containing clouds of average radius
$R_{\star}$ in this region (or a birth radius $\sim0.1R_{\star}$, if
they originate from the surface of the star and expand with the
wind).   The survival times of maser emission
from the clouds scales with their size and is  5 -- $>$10 yr in RSG.

 Most evolved stars which support masers have no detectable rotation
of either the stellar surface or the wind and the presence of a
companion would disrupt masing. The archetype OH/IR stars or H$_{2}$O
envelopes around RSG, for example, appear almost spherical at moderate
resolutions.  This contrasts with the usually axisymmetric shape of planetary
nebulae or into supernova remnants. Early signs of this can be discerned in high-resolution
studies of AGB/RSG stars (e.g. Bowers et al. 1989, Bains et al. 2003;
Murakawa et al. 2003; Etoka \& Diamond 2003), implicating the magnetic
fields detected by e.g. Chapman \& Cohen (1986); Szymczak et
al. (1998, 1999).  We use VLBI and MERLIN observations of OH and other
masers to investigate the CSE of S Per.  This RSG has a distance of
2.3 kpc (Schild 1967) and a stellar velocity $V_{\star}=-38.5$ km
s$^{-1}$ (Diamond et al. 1987). We compare the results for S Per with
those for other stars and with magnetic field models.

\section{Past and present observations of S Per}

   \begin{figure}
   \centering
   \includegraphics[angle=0,width=8cm]{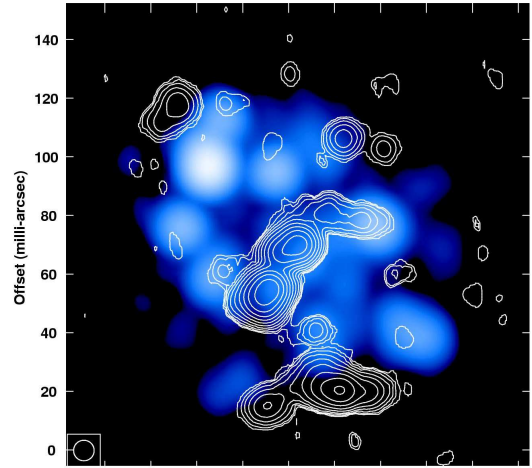}
      \caption{The H$_{2}$O maser emission from S Per (MERLIN 1999) is
      shown in grey overlaid with white contours of OH emission
      observed (VLBI 1997).  The 7-mas VLBI restoring beam is shown
      at bottom left.
         \label{ohh20.fig}
         }
   \end{figure}
   \begin{figure}
   \centering
   \includegraphics[angle=-90,width=7cm]{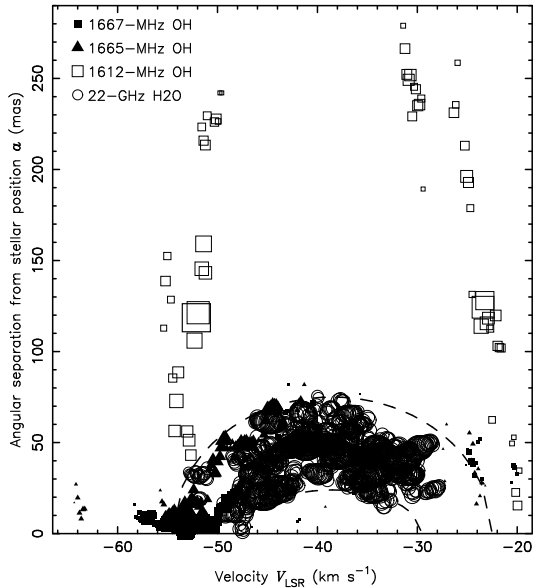}
      \caption{Angular separation of masers from S Per as a
      function of $V_{\rm LSR}$.
         \label{ravel.fig}
         }
   \end{figure}



In 1997 Masheder et al. (1999) observed S Per and VX Sgr with Global
VLBI.
The data were correlated in Socorro
to give a velocity resolution of 0.176 km s$^{-1}$.
We took great care in calibration and editing to remove instrumental
offsets between LL and RR and between the 1665- and 1667-MHz bands, on
all baseline lengths.  We then imaged the brightest channel of the
brightest S Per line to use as a model for iterative phase and
amplitude self-calibration. We applied the resulting solutions to all
the S Per data and then imaged and self-calibrated the other lines
individually.  In this way we were able to retain the correct relative
positions and intensities.  The final data cubes (in LHC and RHC and
in Stokes parameters $I$ and $V$) were made using a 7-mas circular
restoring beam and had a typical noise $\sigma_{\rm rms} \approx 7$
mJy beam$^{-1}$ in quiet channels. We fitted Gaussian components to
measure the properties of each patch of maser emission and grouped
components adjacent in position and velocity to form features thought
to represent physically discrete clumps.  The individual maser
components were mostly spatially resolved, as were the features,
allowing us to measured their angular sizes and line widths.

MERLIN observations made in 1993 (Richards et al. 1999) also showed
 that OH mainline emission appeared to originate at a similar distance
 from the star as the H$_{2}$O masers. In 1999 we made observations in
 full polarization, to produce data cubes in Stokes $I$, $Q$, $U$ and
 $V$ from which we deduced the total linear polarized flux $P$ and
 polarization angle $\chi$.  The individual components and features
 were spatially unresolved.

\label{h20obs}

MERLIN observations of H$_{2}$O masers made in 1994 (Richards et
al. 1999) and 1999 have a similar resolution to the VLBI OH data.  The H$_{2}$O maser shell is consistently well-filled and shows strong
acceleration. At
each epoch the thick shell has a similar size and contains $\sim$100
clumps, about half of which can be cross-identified.  Maser theory
(Cooke \& Elitzur 1985; Yates et al. 1997) suggests that at the inner shell
boundary, 55 au from the star, these have a
number density of $n\sim5\times10^{15}$ which is $50\times$
the wind average at that distance, for a typical temperature
$T\sim1000$ K.
 The integrated maser emission appears almost spherical using a
logarithmic intensity scale but if the brightest components are
selected they are seen to lie in an ellipse elongated NE-SW in 1999.
In 1994 the elongation was more nearly E-W.  Hotspots in an NE-SW
ellipse were also detected in 1998 by Vlemmings et al. 2002 using the
VLBA, see Section~\ref{pol}.

Multi-epoch VLBA monitoring of SiO masers (Ostrowski-Fukuda et
 al. 2002) shows that at some stellar phases the hot-spots trace an
 elongation similar to that of the H$_{2}$O masers.  Thompson \&
 Creech-Eakman (2003) imaged S Per at 2.2$\mu$m and found that the star
 appeared elongated with the minor axis at a similar position angle of
 $40^{\circ}$, suggesting that it is limb-darkened in the direction of
 strongest mass-loss.

\section{Emerging axisymmetries}
\subsection{OH and H$_{2}$O maser distribution}

Fig.~\ref{ohh20.fig} shows the OH mainline and H$_{2}$O maser
distributions, aligned on the centres of expansion, assumed to be the
stellar position. This shows that the OH masers have a very pronounced
elongation NNE-SSW, also seen in 1993. The OH emission is
brightest in between the H$_{2}$O clumps.  The OH 1612-MHz masers have
an angular distribution which is similar in shape to the mainlines but
$>6\times$ greater in extent, covering nearly 1 arcsec N-S (also seen in 1985 observations).  
Fig.~\ref{ravel.fig} shows
the angular separation of masers from the stellar position as a
function of $V_{\rm LSR}$.

These Figs. show that the H$_{2}$O masers form a thick limb-brightened
shell typical of strong acceleration whilst the OH masers are brighter
at the extreme velocities near the front and back caps of the shell
suggesting that they are expanding more steadily. 
For all
species the extreme velocities are found near the centre of emission
and there is no indication of rotation.  Both H$_{2}$O and OH mainline clumps
have an average radius of $r_{\rm c}\sim9$ au with a maximum of almost
30 au.

 The interleaving of OH mainline and H$_2$O masers appears to conflict
 with the classic model in which penetration of interstellar UV forms
 an OH shell outside the H$_{2}$O shell.  It can be explained in the
 context of the high density of the H$_{2}$O clumps mentioned in
 Section~\ref{h20obs}. The dust-gas collision rate is lower at lower
 density.  If the OH comes from surrounding regions at 1/50th the
 density, they are likely to be cooler and less effectively
 accelerated since both heating and expansion are driven by the dust
 absorption of stellar radiation.

   \begin{figure}
   \centering
   \includegraphics[angle=-90,width=7cm]{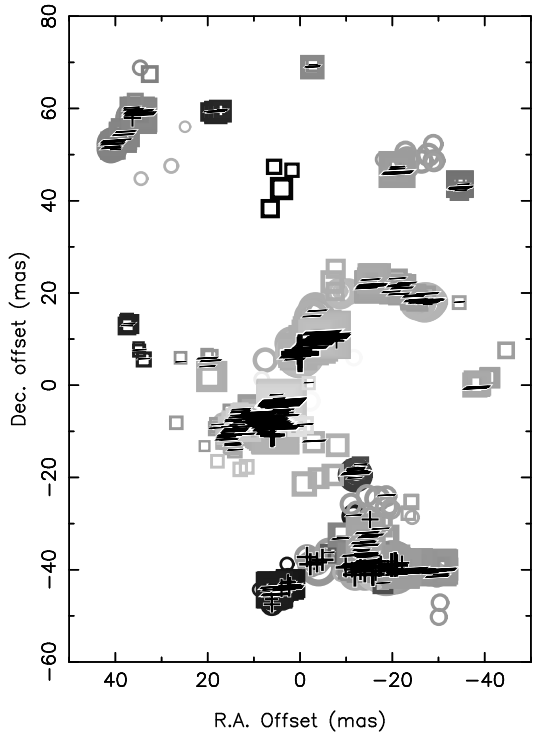}
      \caption{OH masers observed using VLBI. Circles and squares show
      total intensity
      emission at 1665 and 1667 MHz respectively. The darker gray symbols
      show components which are more red-shifted with respect to
      $V_{\star}$. The $+$ and $-$ symbols show the sign of Stokes $V$
      for components with $>25\%$ circular polarization.
         \label{ohv.fig}
         }
   \end{figure}

   \begin{figure}
   \centering
   \includegraphics[angle=-90,width=7cm]{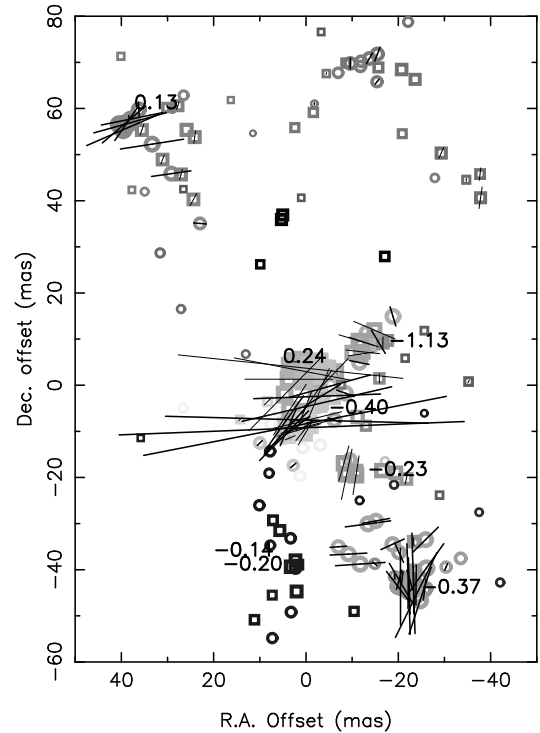}
      \caption{OH masers observed using MERLIN, symbols coded as in
      Fig.~3.  The lines show the linear polarization vectors (thicker
      for 1665 MHz) for components with $>25\%$ linear
      polarization. The numbers are the values of $B_{\parallel}$ in
      $\mu$T inferred from MERLIN circular polarisation measurements.
         \label{ohl.fig}
         }
   \end{figure}

\subsection{Polarization}
\label{pol}
Figs.~\ref{ohv.fig} and~\ref{ohl.fig} show the OH mainline circular
and linear polarised emission.  A 1 km s$^{-1}$
difference between RHC and LHC peaks corresponds to a magnetic field
$B$ of 0.17 and 0.28 $\mu$T at 1665 and 1667 MHz respectively (Davies
1974).  We found 7 possible Zeeman pairs in the EVN data at 1665 MHz
and 2 at 1667 MHz, by looking for features of opposite Stokes $V$
which coincided in position and were the closest matches in velocity.
These had velocity separations (of the Stokes $I$ peaks) of magnitudes
between 0.5--2.8 km s$^{-1}$, giving a magnetic field component
parellel to the line of sight of magnitude $0.1\le
B_{\parallel}\le0.8$ $\mu$T.  Six out of the 9 pairs showed the
magnetic field pointing away from us ($B_{\parallel}>0$).  Where there
were multiple matches we compared  the position and
velocity pattern made by the components within the features. In only one
case an alternative match over 12 km s$^{-1}$ was possible,
giving $B_{\parallel}=3.4$ $\mu$T.  The thermal linewidth of OH at
$T=300-500$ K is 0.9--1.2 km s$^{-1}$; our magnetic field values for
narrower velocity differences may be overestimated (Elitzur 1998).  We
performed a similar analysis of the MERLIN data and found 8 Zeeman
pairs including 6 with $B_{\parallel}<0$. There are also a number of 
maser feature groups where Stokes $V$ is entirely of the same sign.

The mean magnetic field is $B_{\parallel}\approx0.2$ $\mu$T and
$\approx-0.3$ $\mu$T for the EVN and MERLIN data respectively.  There
are some similarities e.g. the most red-shifted pair is seen in all
data with $B_{\parallel}<0$ and the only pair detected in the NNE limb
is at 1665 MHz in both data sets with $B_{\parallel}>0$.

Significant linear polarization is only detected from the near
(blue-shifted) side of the shell and rarely exceeds 50\%, stronger at
1665 MHz.  The dominant direction of polarization vectors $\chi$ is
E-W but there are some anomalies such as in the S where the vectors
change direction by $\sim90^{\circ}$.  The vectors may be parallel or
perpendicular to the component of $B$ in the plane of the sky,
$B_{\perp}$, depending on whether $\pi$ or $\sigma$ maser components
are detected (Elitzur 1998) and also depending on the angle of $B$ to
the line of sight.  The high fraction of circular polarization
relative to linear polarization could suggest that $\sigma$ components
dominate and $B_{\perp}$ is perpendicular to $\chi$, implying that the
magnetic field axis lies along the direction of maser
elongation. However other interpretations are possible (e.g. Deguchi
\& Watson 1986) including Faraday rotation or depolarization.  In any
case it is likely that the $\sim90^{\circ}$ jumps in $\chi$ are caused
by the detection of $\pi$ components or by a small change in the
direction or strength of $B$, not by a full $\sim90^{\circ}$ kink.

Some Faraday rotation is present as shown by the change in $\chi$
across individual features, typically 16$^{\circ}$, which could be
produced by an internal free electron density of $\sim5\times10^7$
m$^{-3}$ across an $\sim18$ au-deep OH masing clump
(a fractional number density of $\sim10^{-6}$). If the OH masers are
distributed in a partly filled spherical shell, the red-shifted
emission travels $\ge100$ au further through the inner CSE which
could, if inhomogenous, convert linear to circular polarization or
produce depolarization averaged over a MERLIN beam.

Vlemmings et al. (2002) used the VLBA to measure Zeeman splitting of
the brightest H$_{2}$O masers.  This is only a small fraction of the
thermal line width as H$_2$O is non-paramagnetic, making analysis
model-dependent. The deduced magnetic field component $B_{\parallel}$
is in the range $15 - 20$ $\mu$T.

\subsection{Biconical outflow model of OH emission}

 The elongation of the 1612-MHz maser shell, which represents material
which left the star several centuries ago, suggests that the OH axis
is stable.  The distribution of  emission suggests a biconical
outflow such as in the models of Zijlstra et al. (2001) .  The
H$_{2}$O masers could  form a complementary distribution with an
equatorially enhanced density surrouding the polar axis of the bicone.
 The slight E-W elongation of the H$_{2}$O masers observed in 1994
appears more compatible with this picture than the more NE-SW
elongation seen in 1999. Projection effects may be confusing the issue
and patience is needed to ensure that the H$_{2}$O elongation is not
chance effects.  Over 6 years, the projected outline of the H$_{2}$O
maser shell of the Mira U Ori has changed apparent orientation,
probably due to chance distribution of bright masers; several decades
would be needed to see if the hotspot distribution around S Per is
systematic.

\section{Polarization interpretation: complications and outlook}

The magnitude and direction of $B$ measured from OH masers shows no
clear pattern nor direction and there are discrepancies between the
MERLIN and EVN data.  This may be due to falsely matched Zeeman pairs
and a more rigorous analysis taking into account single-polarization
features may give a fuller picture. The combination of velocity and
magnetic field gradients can mean that only one sense of polarization
is amplified in a particular clump (Cook 1976).

The average magnetic field strength measured from OH masers is
$\sim0.3$ $\mu$T at a distance from the star of $R\le180$ au; the
largest $B_{\parallel}$ value from a reliable match is $0.8$ $\mu$T.
At our adopted distance of 2.3 kpc the brightest parts of the H$_{2}$O
maser shell are at $R\sim90$ au where Vlemmings et al. (2003) estimate
$B_{\parallel}\sim 15-20$ $\mu$T or $\sim50\times$ the OH value, yet for
a dipole field where $B\propto R^{-3}$ there should be at most a
factor of 8 difference between the field strength in the H$_{2}$O and
OH regions.

The discrepancy could be due to a selection effect; only the strongest
Zeeman splitting of H$_{2}$O masers can be detected but a magnetic
field of even 10 $\mu$T would produce a line split of 59 km s$^{-1}$
for the 1665 MHz OH masers, greater than the total velocity span of
the CSE.  If the magnetic field axis was parallel to the line of sight
the OH masers could lie in the plane of the sky where $B_{\parallel}$
would be very small; however this seems unlikely in view of the
centrally-located extreme velocity OH masers unless the apparent
velocities are due to Zeeman splitting.  

The ratio of thermal to magnetic pressure $\beta=8 \pi \times 10^{-7}n
  k_{\rm B}T/B^2$ is $\sim 0.05$ in the H$_{2}$O maser clumps and
  $\sim2$ in the OH regions, so magnetic pressure is dominant in the
  former and significant in the latter.  It has been suggested
  (including by the present first author) that if the magnetic field
  is a stellar-centred dipole the lighter OH-masing gas could be
  following the dipole axis and/or the dust concentrated in H$_2$O
  clumps could be trapped by the equatorial field lines.  However it
  is hard to see how a static field could shape the wind.  In
  star-forming regions the dust grains spin round the field lines (the
  Whittet-Greenstein mechanism) but in CSEs the grains are the source
  of heating and so such ordered flow would be disrupted by
  thermalisation. Moreover, if this was responsible for concentrating
  the denser dustier clumps in an equatorial belt we would not expect
  to find the strong acceleration which is observed in H$_{2}$O masers
  associated with the clumps.

The origin of a dipole field is also problematic; AGB stars have
already started to accumulate a degenerate core which will ultimately
be exposed as a white dwarf and it has been suggested (Blackman et
al. 2001) that interaction between rotation of the core and convection
cells form a dynamo producing the magnetic field.  This explanation
cannot hold for RSG as they do not become degenerate until the final
catastrophe.  Soaker \& Zoabi (2002) suggest an alternative turbulent
dynamo mechanism which still requires some some internal rotation to
define a symmetry axis. A related possibility is that the pulsation
modes and convection cells in the star create a preferred direction
for denser or less dense mass loss and a stronger or weaker magnetic
field is carried in these directions. Freytag et al. (2002) develop
such a model for $\alpha$ Ori without requiring rotation.  A similar
model for mass-loss involving a shock-compressed field frozen into
clumps at their formation near the stellar surface was suggested by
Hartquist \& Dyson (1997).  The magnetic field strength $B\propto\sqrt
n$. The OH-masing gas is $\sim1/50$ of the density of the H$_{2}$O
clumps at the same distance from the star.  If $n\propto R^{-2}$ then
for OH masers at $R=180$ au, $B$ would be $\sim1/15$ of its value for
H$_{2}$O masers at 90 au, closer to the observed strengths.
 
 The supergiant VX Sgr does
appear to have an ordered magnetic field measured from OH 1612 MHz
masers (Szymczak \& Cohen 1997; Szymczak et al. 2001), the axis of
which defines a biconical region containing a lower density of
H$_{2}$O masers (Murakawa et al. 2003). The full analysis of OH
mainline masers observed by the EVN/Global VLBI and by MERLIN will
confirm or modify this model.  The magnetic field configuration
appears to evolve with time; OH masers around NML Cyg show an old shell with tangential
polarization vectors surrounding a more recent bipolar outflow
parallel to radial polarization vectors (Etoka \& Diamond 2004).
Full 3-D modelling of S Per is needed, if possible including
further epochs of observation to measure proper motions and examine
the persistence of the magnetic field. 

\end{document}

%% file: page.tex
\setcounter{page}{209}